\documentclass[12pt]{iopart}
\usepackage{graphicx}
\usepackage{color}
\usepackage{dcolumn}
\usepackage{bm}
\newcommand{\p}{\\[2ex]}

\newcommand{\Fig}[1]{Figure~\ref{#1}}
\newlength{\figwidth}
\newcommand{\tabl}[1]{table~\ref{#1}}

\newlength{\hw}
\newlength{\vvp}
\newlength{\minusspace}
\newcommand{\msp}{\hspace{\minusspace}}
\newlength{\zerospace}
\newcommand{\sstrut}{\rule[-1ex]{0ex}{3ex}}
\newcommand{\degree}{\ensuremath{^\circ}}
\newcommand{\mub}{\ensuremath{\mu_{\mathrm B}}}

\newcommand{\kv}{\ensuremath{\textbf{k}}}
\newcommand{\ustar}{^{*}}

\newcommand{\N}{{\ensuremath{N}}}

\newcommand{\Mi}[1]{{\ensuremath{\textbf{M}_{#1}}}}

\newcommand{\Qv}{{\ensuremath{{\bf M}_{\perp}}}}
\newcommand{\Qvs}{\ensuremath{{\bf M}_{\perp}\ustar}}

\newcommand{\Qi}[1]{{\ensuremath{M_{\perp#1}}}}

\newcommand{\Q}{{\ensuremath{M_{\perp}}}}

\newcommand{\hfo}{HoFeO$_3$}
\newcommand{\hf}{\ensuremath{\frac12}}
\newcommand{\tqr}{\ensuremath{\frac14}}
\newcommand{\qr}{\ensuremath{\frac34}}

 \mathchardef\mhyphen="2D
\begin{document}
\title[Polarised Neutron diffraction study of HoFeO$_3$]{Single crystal polarized neutron diffraction study of the magnetic structure of HoFeO$_3$}
\author{T. Chatterji, A. Stunault and P.J. Brown}
\address{Institut Laue-Langevin, 6 rue Joules Horowitz, BP 156, 38042 Grenoble Cedex 9, France\\
 }
\date{\today}
\begin{abstract}
Polarised neutron diffraction measurements have been made on HoFeO$_3$ single crystals magnetised in both the [001] and [100] directions ($Pbnm$ setting). The polarisation dependencies of Bragg reflection intensities were measured both with a high field of H = 9 T parallel to [001] at T = 70 K and with the lower field H = 0.5 T parallel to [100] at T = 5, 15, 25~K. A Fourier projection of magnetization induced parallel to [001], made using the $hk0$ reflections measured in  9~T, indicates that almost all of it is due to  alignment of Ho moments. Further analysis of the asymmetries of general reflections in these data showed that although, at 70~K, 9~T applied parallel to [001] hardly perturbs the antiferromagnetic order of the Fe sublattices, it induces  significant  antiferromagnetic order of the Ho sublattices in the $x\mhyphen y$ plane, with the antiferromagnetic components of moment having
the same order of magnitude as the induced ferromagnetic ones.  Strong intensity asymmetries measured
in the low temperature $\Gamma_2$ structure with a lower field, 0.5 T $\parallel$ [100] allowed the variation of the ordered components of the Ho and Fe moments  to be followed. Their absolute orientations, in the 180\degree\ domain stabilised by the field were determined relative to the distorted perovskite structure,. This relationship fixes the sign of the Dzyalshinski-Moriya (D-M) interaction which leads to the weak ferromagnetism. Our results indicate that the combination of strong y-axis anisotropy of the Ho moments and Ho-Fe exchange interactions breaks the centrosymmetry of the structure and could lead to ferroelectric polarization.    
\end{abstract}
\noindent{\it \hfo, polarized neutrons,weak-antiferromagnetism}/ 
\maketitle 

\section{Introduction}
The family of rare-earth orthoferrites RFeO$_3$ (R = rare-earth element) contains a large number of compounds with 
interesting magnetic and other physical properties 
\cite{white69,belov76,belov79,treves65,gyorgy68,sirvardiere69,hornreich73,yamaguchi74}. These compounds have unusually
 high N\'eel 
temperatures, 600-700 K, below which the Fe moments order with basically antiferromagnetic 
structures which also often exhibit  weak ferromagnetism.  Recently  interest has focussed on the possiblity of 
multiferroic properties arising from the complex interactions between 3d and 4f moments 
\cite{tokunaga08,tokunaga09,lee11,kuo14}. If such interactions lead to the development of a second ferroic order, such as 
ferroelectricity, at relatively high temperature, the compounds could have useful device applications. Previous studies 
of the rare earth  orthoferrites have shown that they undergo interesting spin reorientation transitions at temperatures 
below about 50 K. On further cooling the magnetic rare earth ions, which at higher temperature are  paramagnetic, or 
weakly polarized by the molecular field of the ordered Fe ions, may also become magnetically ordered. Complicated 
magnetic properties arise from the multiplicity  of different exchange interactions. The three principal exchange 
interactions have strengths decreasing in the order Fe-Fe, R-Fe and R-R. At high temperature the structure is determined 
by Fe-Fe interactions but as the temperature is lowered magnetic transitions may be triggered by the increasing importance 
of the Fe-R interaction. Finally at very low temperatures the R-R interaction dominates the fully ordered magnetic 
structure. 

    In a recent paper \cite{chatterji17} we reported the results of a single-crystal unpolarized neutron diffraction 
study of the temperature evolution of the magnetic structure of HoFeO$_3$. This detailed study confirmed that HoFeO$_3$ 
orders at a very high temperature ($T_N \approx 700$ K) with a $\Gamma_4$ structure \cite{sirvardiere69}.
This high 
temperature phase is weakly ferromagnetic in the [100] direction and is stable in the temperature range from $T_N \approx 700$ - 55~K  
At 55~K there is a abrupt spin re-orientatiion transition to a $\Gamma_1$ structure and the weak ferromagnetism is lost. 
Below about  35 K a previously unreported and non-abrupt reorientation transition to the $\Gamma_2$ structure begins. 
The $\Gamma_2$ structure is again weakly ferromagnetic but with x as the ferromagnetic axis. This $\Gamma_2$ structure 
undergoes two further changes as the degree of Ho order increases on cooling below 35 K. At $T\approx 20$ K and again at 
$T \approx 10$ K the relative directions of the ferromagnetic component of Fe and Ho change sign. 
\p
 Since the magnetic structures of both the high and low temperature structures of \hfo\ exhibit weak ferromagnetism
it is of some interest to study their response to applied magnetic fields. Polarised neutron Bragg scattering is the
tool of choice for such a study as it exploits the coupling between the neutron polarization and that induced in the sample and can  yield  directional information not easily obtainable by other methods. The present paper reports the results
we have obtained for \hfo\ using this technique.

\section{Polarised neutron Bragg scattering in weak ferromagnets}
The magnetic polarization induced in a crystal by an applied field gives rise to a magnetic structure factor whose
phase, relative to that of the nuclear structure factor, depends on the magnetization direction, and is reversed
if the magnetization is reversed. The intensity of Bragg scattering of polarised neutrons likewise depends on the relative
orientations of the neutron polarization and the magmetic structure factor.  
The resulting polarization dependence of  intensity 
(polarised neutron intensity asymmetry) is used in classical polarised neutron experiments to measure magnetization 
distributions. In cases such as that of \hfo\ in which some  of the magnetic species have significant anisotropy there
may be components of magnetic moments perpendicular to the applied field and ordered either ferromagnetically or 
antiferromagnetically. These also contribute to the polarised neutron asymmetries and their magnitudes can be determined from polarised 
neutron experiments \cite{gukasov2002,brown2005}
\p
The polarised neutron asymmetries may also be used to determine the sign of the D-M interaction. 
To exhibit (D-M) weak ferromagnetism  a material must have a magnetic structure with zero propagation
vector and the magnetic symmetry group must contain an element combining rotation and translation which relates atoms with nearly antiparallel spins. Applying a magnetic field to such a structure breaks  the degeneracy of the two 180\degree\ domains (those in which all the antiferromagnetic components are reversed), favouring the one with the ferromagnetic component parallel to the field. The magnetic structure factors of the two 180\degree\ domains  differ in phase by 
180\degree\ whereas their nuclear structure factors are the same. Unbalancing the domain populations therefore leads to
a polarised neutron intensity asymmetry whose sign depends on that of the D-M interaction (the absolute direction in which the magnetic moments are canted with respect to their atomic environment)\cite{brown1981,brown2011}

\section{Experimental methods}
\subsection{Crystal growth and characterization}
The HoFeO$_3$ single crystals used in the present experiment were grown using flux method by S.B.N. Barilo and D.I. Zhigunov in Minsk, Belarus. The crystals were of very good quality giving sharp diffraction peaks. Magnetization and specific heat measurements on a small 122 mg single crystal of HoFeO$_3$ cut out of the larger single crystal have been reported by us \cite{chatterji17}  and these results are in agreement with the results reported by other authors \cite{shao11}. We have also previously reported  \cite{chatterji17} the temperature evolution of the magnetic structure  of these same \hfo\ crystals using unpolarized neutron diffraction.
\subsection{The polarized neutron diffraction}
Polarized neutron diffraction measurements were made on these HoFeO$_3$ single crystals using the spin polarized difractometer D3 of the Institut Laue-Langevin in Grenoble. The crystals were magnetised and cooled using a cryomagnet delivering a maximum vertical field of 9 T. The polarised neutron intensity asymmetries of accessible Bragg reflections  
were measured in both 9~T and in the lowest field retaining good polarization  viz 0.5~T. The high field measurements were made at 70~K with the crystal magnetised parallel to [001] (Pbnm setting) and the low field  measurements  at selected temperatures 
in the range 5-60~K with the \hfo\ crystal magnetised parallel to [100]. 

The magnetic structure models, the nuclear structure parameters and the extinction model used in analysing the data were those determined from the unpolarised single crystal measurements   \cite{chatterji17} made at the FRM II reactor, Garching. The space groups used are  Pbnm, and its subgroup Pbm2$_1$. The labels given to the different magnetic sublattices and the relative orientations of 
the components of their magnetic moments in the $\Gamma_4$ and $\Gamma_2$ structures are summarised in \tabl{strucs}
\begin{table}[h]
\caption{\label{strucs}Relative orientations of the x,y and z components of moments on the 4 Fe and 4 Ho sublattices in 
the structure of \hfo\ for the $\Gamma_4$ (Pb'n'm and $\Gamma_2$ (Pbn'm') magnetic phases. For Ho x=0.9186 y=0.0685.}
\setlength{\hw}{-1.2ex}
\begin{indented}
\item \begin{tabular} {lccccccccccc}
\br
&&&&&\multicolumn{6}{c}{Phase}\\
Atom &\multicolumn{3}{c}{Position}&\quad&
\multicolumn{3}{c}{$\Gamma_4$ T$>55$K}&&\multicolumn{3}{c}{$\Gamma_2$ T$<35$K}\\
 & x &  y & z&&$M_x$ & $M_y$ &$M_z$&& $M_x$ &$M_y$ & $M_z$\\
\mr
Fe1 &  0  & \hf \sstrut&   0 \\[\hw]
&&&&&+&+&+&&+&+&+\\[\hw]
Ho1 & x & y & \qr\sstrut\\[1ex]
Fe2 & \hf &   0 & 0 &\sstrut\\[\hw]
&&&&&$-$&+&+&&+&$-$&$-$\\[\hw]
Ho2 & \hf$-$x  & \hf+y & \sstrut\qr\\[1ex]
Fe3 & \hf &   0 & \hf &\sstrut\\[\hw]
&&&&&+&$-$&+&&+&$-$&+\\[\hw]
Ho3 & \hf+x & \hf$-$y &\tqr \sstrut\\[1ex]
Fe4  & 0   & \hf & \hf \sstrut\\[\hw]
&&&&&$-$&$-$&+&&+&+&$-$\\[\hw]
Ho4 & $-$x & $-$y & \tqr \\
\br
\end{tabular}
\end{indented}
\end{table}

\section {[001]Orientation, 9T applied field at 70~K}
At 70~K the crystal has the $\Gamma_4$ structure in which only the Fe moments are ordered. Applying
magnetic field parallel to [001], which is the axis of weak ferromagnetism, should initially sweep out 180\degree\
 domain walls so that the weak ferromagnetic moment is aligned parallel to the magnetic field throughout the crystal.
 Further increase in field will result in progressive paramagnetic alignment of Ho moments and canting of 
 the antiferromagnetic components of Fe moments into the field direction. 
 
\begin{figure}[h]
\begin{center}
  \resizebox{0.7\textwidth}{!}{\includegraphics{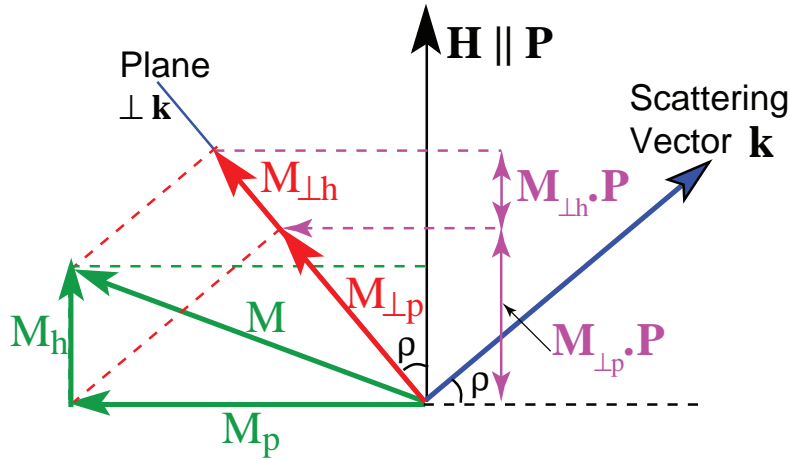}}
\caption{Vector diagram illustrating the relative directions of vectors in equations (1), (2) and (3)
with $\phi=0$.}
\label{projection}
\end{center}
\end{figure}

\begin{figure}[h]
\begin{center}
   \resizebox{0.6\textwidth}{!}{\includegraphics{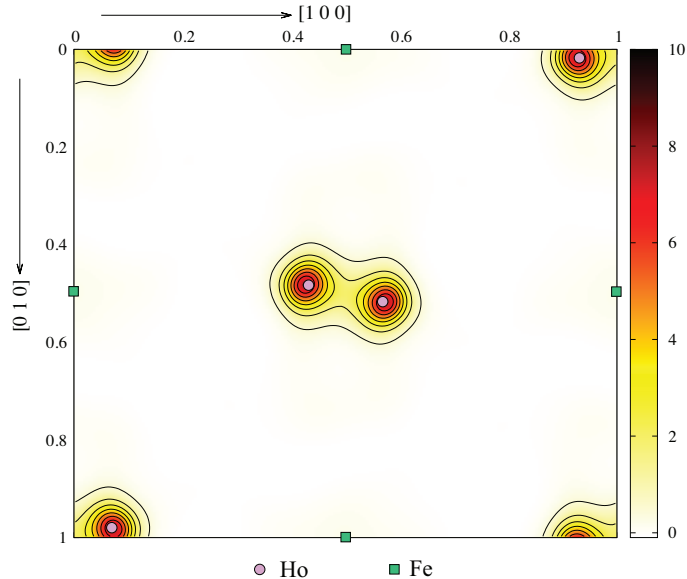}}
\caption{Maximum entropy reconstruction of the magnetization of \hfo\ aligned parallel to [001] by a 9~T field applied parallel to [001] at 7~K. The contours are drawn at intervals of 1.0~\mub\AA$^{-2}$ starting at 0.5.}
\label{maxe}
\end{center}
\end{figure}

 The quantity measured in the polarised neutron experiment is the intensity asymmetry
$A=(I^+ - I^-)/(I^+ + I^-)$. In this equation $I^+$ and $I^-$ are the intensities of neutrons scattered  
with polarization parallel (+) and antiparallel (-) to the field direction. With the normal beam geometry 
 of D3, in which the field is parallel to the crystal rotation axis and perpendicular to the incident beam, the 
cross-sections for Bragg  scattering  are
 \begin{eqnarray}
I^{\pm}\propto |\N|^2& + &|\Q|^2 \pm2\Re(\N\Qvs)\cdot\hat{\bf P}\\
\mbox{giving}\ 
A& = &\frac{2\Re(\N\Qvs)\cdot\hat{\bf P}}{|\N|^2 + |\Qv|^2 }
\end{eqnarray}
where \N\ is the nuclear structure factor,  \Qv\ the magnetic interaction vector of the reflection and
$\hat{\bf P}$ a unit vector parallel to both the polarization and the magnetic field. In centrosymmetric \hfo\ the structure factors
are real and the aligned magnetization can be separated into components $M_{h}$ and \Mi{p} parallel and 
perpendicular to the field. The relative orientations of these vectors for a reflection whose scattering vector \kv\ is inclined at angles $\rho$ to \textbf{P} and $\phi$ to \Mi{p}\ in the plane perpendicular to \textbf{P}, 
are shown in figure~\ref{projection}. The magnetic interaction vectors, obtained by projecting the magnetic structure factors ${\bf M}$ onto the plane perpendicular to  \kv\,
have the form 
\begin{eqnarray}
\Qi{h}&=&M_{h}\sin\rho\nonumber\\ \mbox{and}\qquad \Qi{p}&=&M_{p}\cos\rho\sin\phi
\end{eqnarray}
With an [001] crystal orientation $\rho=\pi/2$ for all reflections $hk0$, so the asymmetries measured for 
these reflections depend  directly on the magnetic moment $M_{h}$ aligned by the field. However these data cannot be used to 
determine its distribution between the Fe and Ho atoms directly, using Fourier techniques, when the contributions to $|
\Q|^2$ of any antiferromagetically aligned moments are not known. Furthermore, any inaccuracy in aligning the chosen crystallographic axis parallel to the field direction allows antiferromagnetic components to contribute to the $hk0$ asymmetries. Nevertheless a preliminary analysis of the $hk0$ reflection asymmetries was made assuming that these antiferromagnetic contributions  could be neglected. This yielded the Fourier projection 
figure~\ref{maxe} which strongly suggests that almost all the 
magnetization is due to alignment of Ho moments.
  In further analysis the measured asymmetries were 
compared to those predicted by several different models for the magnetized structure. In the centrosymmetric space group  Pbnm the Ho atoms lie on the mirror planes perpendicular to $ [001]$, which constrains their moments to lie either parallel or perpendicular to that direction. It was not found possible to obtain a reasonable fit to the asymmetries with models retaining this constraint. The best agreement was found with
models in which the aligned moments on the Ho sublattices follow the same $\Gamma_4$ coupling scheme as the Fe 
atoms and that they have significant ordered components of magnetization parallel to both [010] and [001]. These models imply the lower space group symmetry, Pbn2${_1}$. Least squares  fits of the components of magnetization of Fe and Ho to the asymmetries calculated for this coupling scheme and space group gave the results shown in table~\ref{paramu}.
Initially the the $\langle j_0\rangle$ (spin only) form factor for Ho$^{3+}$ was used to model the Ho magnetic scattering, however the 
fit was  significantly improved by including a $\langle j_2\rangle$ (orbital) term. In the dipole approximation form factor for
magnetic scattering is given by:
\begin{equation} f(K) = (M_S + M_L),\langle j_0(k)\rangle +  M_L\langle j_2(k)\rangle \end{equation}
where $M_S$ and $M_L$ are the spin and orbital moments and $\langle j_0(k)\rangle$ and $\langle j_2(k)\rangle$ the radial integrals of order 0 and 2 for the scattering ion \cite{marshall71}
\begin{table}[h]
\caption{\label{paramu}Components of magnetization on Ho and Fe in \hfo\ at 70~K aligned  by 9~T applied parallel to z.}
\setlength{\hw}{-1.2ex}
\begin{indented} \item
\begin{tabular}{clllll}
\br
&&\multicolumn{3}{c}{Magnetization (\mub)}\\[\hw]
Atom &Model&&&&$R_{\mbox{\scriptsize{cryst}}}$\\[\hw]
 &&M$_x$  &M$_y$  &M$_z$\\
\mr
Fe&Spin & 4.5994(8)   & -0.04(9)  & 0.060(14)\\[\hw]
&&&&&20.2\\[\hw]
 Ho&Spin &0.0(5)  &  1.58(10)  &   1.70(9)\\[1ex]
Fe&Spin &4.597(2)  &  -0.12(8)  & 0.114(10)\\
Ho& Spin&0.6(2)  &  0.75(14)  &    0.97(9)& 8.7\\
Ho& Orbital&  0.20(9)  &    0.41(6)   &   0.53(6)\\
\br
\end{tabular}
\end{indented}
\end{table}%

The results in table~\ref{paramu} suggest the magneto-crystalline anisotropy of Ho favours the y orientation and that even with 9~T applied parallel to  [001] almost half of the aligned Ho moments are antiferromagnetically arranged in the $x\mhyphen y$
plane. 

\section {Polarization dependent intensities in the $\Gamma_2$ phase with H=0.5 T $\parallel [100]$}
At temperatures below 35~K the $\Gamma_2$ structure is stable with the principal components of the Fe and Ho moments
parallel to  [001] and [010] respectively. It exhibits weak ferromagnetism parallel to [100] \cite{chatterji17}. In the temperature 
range 35-2~K the ordered Ho moment increases from less than 1~\mub\ to almost 8~\mub\ at 2~K. Measurements of  
polarised neutron intensity asymmetries of a set of 276 reflections were made at 20, 15, and 5~K to investigate how 
the changing strengths of the Fe-Fe, Fe-Ho and Ho-Ho interactions influence details of the structure. For these 
measurements the x axis of the crystal was nearly parallel to the direction of the 0.5~T magnetic field. With the 9~T cryomagnet a minimum field of 0.5~T is
required to maintain the neutron polarization; it  also ensured that the weak ferromagnetic moment was fully aligned.

The magnetic parameters of the ${\Gamma_2}$ structure were fitted by a least squares procedure to the measured 
asymmetries  at each of the measurement temperatures. The magnetic group $Pbn'2_1'$ was used and both 
$\langle j_0\rangle$ and $\langle j_2\rangle$ contributions were included in the Ho form factor.

\begin{table}[h]
\caption{\label{xaxe}Components of magnetic moment on Ho and Fe  atoms in the antiferromagnetic 
$\Gamma_2$ structure of \hfo\ measured with 0.5~T $\parallel [100]$
at 25, 15 and 5 K.\\[-1ex]}
\begin{indented} \item
\setlength{\hw}{-1.2ex}
\begin{tabular}{cllllc}
\br
&&\multicolumn{3}{c}{Magnetic Moment (\mub)}\\[\hw]
T(K)&Atom &&&&$R_{\mbox{\scriptsize{cryst}}}$\\[\hw]
 &&\multicolumn{1}{c}{M$_x$}  &\multicolumn{1}{c}{M$_y$}&\multicolumn{1}{c}{M$_z$}\\
\mr
   &Ho(S)   &   $-$0.91(5)   &  $-$2.13(10) &  \msp0.57(3)\\
25 &Ho(L)  &   $-$0.19(4)   &  $-$0.45(10) &  \msp0.12(3)&11\\
   &Fe     &   $-$0.01(6)   &  \msp0.12(4) &  \msp4.60(1)\\[1ex]
 
   &Ho(S) &  $-$1.36(9)   & $-$3.2(2)    & \msp0.58(5)\\
15 &Ho(L) &  $-$0.27(8)   & $-$0.6(2)    & \msp0.11(3)&11\\
   &Fe    &   \msp0.01(1) &  \msp0.20(6) & \msp4.60(1)\\[1ex]
 
   &Ho(S)  &   $-$2.1(3)  &  $-$5.4(6)  & \msp0.80(11)\\
 5 &Ho(L)  &  \msp0.2(2)  & \msp0.6(6)  &  $-$0.08(9)&14\\
   &Fe     &  \msp0.11(6) & \msp0.8(2)  & \msp4.54(4)\\
\br
\end{tabular}
\end{indented}
\end{table}%

The results obtained from refinements,using the measured
polarised neutron asymmetries as data, are given in \tabl{xaxe}. These results confirm that the 
easy direction of the Ho  moment is parallel to y. 

\section{Discussion}
\subsection{The sign of the antisymmetric coupling in \hfo}
The sign of the antisymmetric coupling which leads to weak ferromagnetism in \hfo\ can be inferred from the magnetic 
moments given in table~\ref{xaxe} since the listed components are those on the representative atoms (Fe at 0\hf 0 and
Ho at (0.9186, 0.0685, \qr).  It can also be determined directly from the signs of the asymmetries. In relatively low 
fields the antiferromagnetic
components of moment are much larger than the weak ferromagnetic ones and so dominate the interaction vectors. These 
larger antiferromagnetic components are oppositely oriented  in the two 180\degree\ domains and their phases relative to 
those of the nuclear structure factors determine the signs of the asymmetries of the corresponding reflections. In
the present experiment the axis parallel to the magnetic field and polarization direction was not exactly [100] but
[-0.9782,-0.1907,0.0823] allowing both the $y$ and $z$ antiferromagnetic components  to contribute to the asymmetries  in 
reflections with $h \ne 0$. For example the asymmetry measured in the 121 reflection at 25~K was -0.263(8); its nuclear 
structure factor calculated with the atomic positions given in table~\ref{strucs} is -2.1971 ($10^{-12}$~cm) and the 
real part of the magnetic 
interaction vector for the 25~K $\Gamma_2$ structure (table~\ref{xaxe}) has components x=-0.8730,y=-0.4709, z=2.5561  ($10^{-12}$~cm).
For the other 180\degree\ domain in which the $y$ and $z$ components of moments are reversed the components of the 
interaction vector are x=0.7981,   y=0.5059, z=-2.5433. The product $\Re({\bf P}\cdot\Qv)$ which leads to the asymmetry 
is negative for the first configuration and positive for the second. The weak ferromagnetic moment which has the same sign for both 180\degree\ domains is responsible for
the small differences between the absolute values of the components. 

\begin{figure}[htbp]
\begin{center}
 \resizebox{.6\textwidth}{!}{\includegraphics{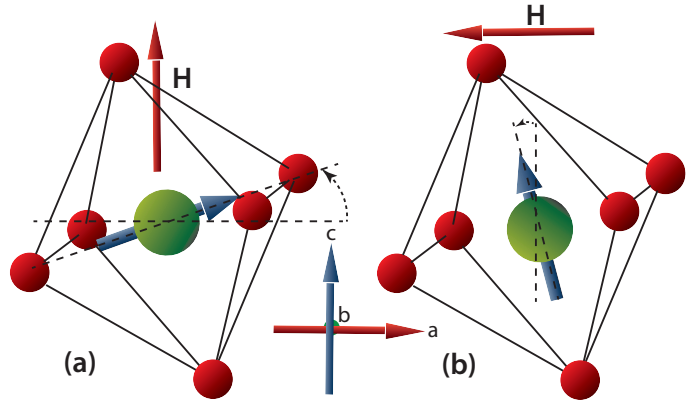}}
\caption{Schematic drawing showing the orientation of the moment on the Fe atom at (0 \hf 0) in \hfo\ relative to its coordinating octahedron of O atoms. (a) in the $\Gamma_4$ structure at 60~K and (b) In the $\Gamma_2$ structure at 25~K. For clarity the direction of rotation of the Fe moment giving rise to weak ferromagnetism is indicated by the dashed lines and is exagerated.}
\label{octs}
\end{center}
\end{figure}

In the perovskite structure the transition metal atoms are coordinated by octahedra of O atoms which share
vertices. In the ideal structure the axes of the octahedra are parallel to the crystal axes but in the distorted perovskite structure of \hfo\ they are tilted and twisted so that their axes lie in directions: 
 x=[0.84   0.53   0.15],  y=[0.53  -0.84   0.15], z=[-0.40 0.14 -0.91],
 \Fig{octs} shows schematically the orientation of the moment on the Fe atom at (0,\hf,0) relative to its  coordinating octahedron  (a) at 70~K in a field of 9~T parallel to [001] and (b) at 25~K in a field of 0.5~T nearly parallel to 
[$\bar1$00]] In both cases the tilt of the Fe moment towards the field direction is in the same direction as the tilt 
of its coordinating octahedron. 
At 25~K the ferromagnetic component of the Fe moment is very small and the preferred domain must be determined by the Fe-Ho interactions.
In the low temperature data the contribution of Fe to the weak ferromagnetism is very small relative to that of Ho and and the precision with which it can be determined is limited both by this and by interference from antiferromagnetic
components due to imperfect crystal alignment. Nevertheless the variation of the ferromagnetic (x) component of Fe moment  between 25 and 5~K,
although hardly significant, is  in accord with the changes in reflection intensities in this temperature range found in zero field measurements \cite{chatterji17}. From these it was deduced that the relative directions of the antiferromagnetic Ho and Fe moments change sign at around 15~K. The resultant direction of magnetization is determined by the larger 
 Ho contribution so that when the relative directions of the moments change sign  
the weaker ferromagnetic moments of the Fe atoms align in the opposite direction to the magnetising field
leading to reversal of the Fe $x$ component.
 
\subsection{Symmetry}
The foregoing analysis has been carried out assuming  the atomic positions of the nuclear space group Pbnm.
However, maintaining this space group for the magnetic structure, constrains the Ho moments to lie either
parallel or perpendicular to [001]. The magnetic fields applied in the present experiments can break the
mirror symmetry of the planes on which the Ho atoms lie. The consequent loss of centrosymmetry permits
the oxygen octahedra to acquire a polar character allowing ferroelectricity. The results of the measurements
in the $\Gamma_4$ phase at 60~K show that the Ho moments have very high magneto-crystalline anisotropy so that
a field applied parallel to  [001] induces antiferromagnetic ordering parallel to [010],in addition to ferromagnetic
alignment along  [001]; a configuration not consistent with centrosymmetry. In the low temperature $\Gamma_2$ phase
there is a small but significant component of Ho moment parallel to [001], again indicating a loss  of
centrosymmetry presumably due  to exchange interaction with the Fe moments which are aligned parallel to $z$ in this phase.
\section{Conclusion}
Polarized neutron diffraction measurements of the intensity asymmetries induced  by applying magnetic fields can in principle give important information (1) about the magnetic anisotropy of some of the magnetic species and (2), can  be used to determine the sign of the D-M interaction. The present measurements of intensity asymmetries from HoFeO$_3$ with the $ k = (0,0,0)$ magnetic structures $\Gamma_4$, and $\Gamma_2$ type have enabled us to extract both types of information. The results of our experiment shows that strong magnetocrystalline anisotropy of Ho favours the y orientation and that with even H= 9 T applied field parallel to [001] almost half of the ordered Ho moment is not aligned parallel to the field but antiferromagnetically arranged in the $x\mhyphen y$ plane. In addition the sign of the asymmetric coupling can be inferred from the magnetic moment components given in table III since the listed moments are those on the representative atom (Fe at 0,1/2,0 and Ho a0.0685, 0.9186, 3/4). It can also be determined more directly from the signs of the intensity asymmetries. Another important conclusion (3) from the present investigation is that, due to the Ho anisotropy the centrosymmetry 
of the structure is broken: in the $\Gamma_4$ phase by a magnetic field parallel to [001] and in the low temperature $\Gamma_2$ phase by its interaction with ordered Fe moments. The loss of centrosymmetry of the Fe site, which this implies, would allow its coordinating octahedron of oxygen to become polarised, and consequently allow ferroelectricity. At temperatures below the Fe ordering temperature, even with no applied field, the Ho are subject to the  internal field  due to the ordered Fe moments. It is conceivable that, due to the Ho anisotropy, the internal field could induce moments on Ho which will break the centrosymmetry. So there is a possibility that HoFeO$_3$ develop ferroelectricity at low temperatures.



\end{document}